\documentstyle[pre,aps]{revtex}

\begin{document} 
\title{ Random Energy Model at Complex Temperatures.}

\author{D.B. Saakian}
\address{Yerevan Physics Institute,
Alikhanian Brothers St. 2, Yerevan 375036, Armenia\\ and \\
Universidad de La Frontera, Departamento de Cienas Fisicas, Casilla
54-D, Temuco, Chile.
}

\maketitle

\begin{abstract}
The complete phase diagram of Random Energy Model (REM) is obtained for
complex temperatures
using the method proposed by Derrida. We find the density of zeroes for 
statistical sum. Then the method is applied to Generalized Random Energy
Model (GREM). This allows us to propose new analytical
method for investigating zeroes of statistical sum for finite-dimensional
systems.
\end{abstract}

\vspace{1cm}

Random Energy Model (REM) [1] is one of the most popular contemporary
models of 
statistical mechanics. Besides its direct applications in spin-glasses the
model has a wide
range of employments in very different areas of the modern theoretical
physics and biophysics.
Recently, several new, and more intimate applications have been founded 
in the theories of well-developed turbulence [2], and strings [3-5].
On the other hand, it is well-known that the model allows to consider the
fundamental problems 
of information theory in the language of statistical mechanics. In
particular, problems connected
with transmission of information through a noisy channel can be formulated
and effectively solved 
in the language of REM [6,7]. This astonishing range of different
applications cannot be 
accidental one:   
It seems, that REM is a paradigm for a complex system physics, or
at least contains all essential ingredients of that physics. In other
words,
the nature sings his sacral songs using REM's language.

The thermodynamical structure of REM has simple structure but at the same
contains all essential
ingradients of phase transitions \cite{B1,B11,B12}: There is well defined
critical temperature
$T_c$ which
can be obtained from clear and semi-intuitive physical reasons. For
$T<T_c$ the system is frozen in
the spin-glass phase with one level of replica symmetry breaking. Besides
this rough
thermodynamical phase transition there are some "mild", mesoscopic phase
transitions in the domain
$(\tilde{T}, T_c)$, $\tilde{T}> T_c$ \cite{B1,B12}. The temperature
$\tilde{T}$ strongly depends
from the
concrete form of energy's distribution; it can be infinite for the
gaussian distribution \cite{B1} 
but it has
some finite value for REM generated by more physical dilute hamiltonians
\cite{B12}.
This rich and interesting spectra of phase transitions waits its complete
and exhaustive
investigation, and we hope that the present work will shed some light to
this question also.

It is well-known that structure and properties of phase transitions can be
investigated through
analytical
properties of thermodynamical quantities in the complex plane of
temperature or/and magnetic field
\cite{grif}\cite{zuber}\cite{armen&saco}.
Indeed, if a phase transition is assosiated with singular behavior of
thermodynamic potential in
the thermodynamic limit (in our case it is free energy or statistical
sum), then by consideration
of its analytical properties in the complex plane the important physical
information can be
obtained. The method was proposed by Yang and Lee \cite{yang} (for the 
case of complex magnetic
fields), and Fisher \cite{fisher} (for the case of complex temperatures).
It has large variety of different applications in statistical
physics; furthermore it 
is one one of the most exact and powerful methods for estimating critical
indices and constructing 
phase diagrams. In particular, the method is applied for investigation of 
phase transitions in disordered or strongly frustrated systems because
sometimes all other
methods are not very
informative or hard for realization. Resently with help of this method a
dense
domain of phase transitions has been obtained in a non-disordered but
fully-frustrated model
\cite{armen&saco}.
 
REM has been investigated in complex field/temperature plane [8,10].
The goal of the present paper is to continue this program, and to derive
the complete phase diagram. Furthermore,
we propose a new method for investigating analytical properties of
thermodynamical quantities of finite-dimensional systems.
There is another, perhaps more fundamental motivation for the present
investigation:
it was obtained resently that REM at complex temperatures is closely
connected with strings \cite{B4}\cite{B5}.

In disordered systems there are averaging with the canonical Gibbs 
distribution at
given temperature, and averaging by frozen disorder. In REM and related
models
energy levels are considered as random quantities, and the distribution
is derived from a concrete "microscopic" Hamiltonian
\cite{B1}\cite{B2}.
\begin{equation}
\label{ee1}
H=-\sum_{1\le i_1<i_2..<i_p\le N}j_{i_1..i_p}s_{i_1}..s_{i_p}
\end{equation}
Here $s_i$ are $\pm 1$ spins, $j_{i_1..i_p}$ are quenched couplings with 
normal distribution and a proper scale
\begin{equation}
\label{ee2}
\rho(j_{i_1..i_p})=\sqrt{\frac{2C^N_p}{N\pi}}\exp{-2C^N_pj_{i_1..i_p}^2/N},
\end{equation}
where $C^N_p=N!/p!(N-p)!$.
It is possible to prove, that in the limit of large $p$ all energy levels 
are distributed independently with probability

\begin{equation}
\label{ee3}
P(E)=\sqrt{\frac{1}{N\pi}}\exp (-E^2/N)
\end{equation}
Let us define partition function and free energy. 
Consideration of complex temperatures is some tool for investigating 
analitical properties of thermodynamical quantities. In particular,
the usual definitions of statistical physics should be adapted to this 
purpose.
Partition function is defined as usually:
$$
Z=\sum_\alpha e^{-\beta E_\alpha}
$$
Further, at the thermodynamical limit different expressions  like
\begin{equation}
\label{ee4}
<\ln\mid Z\mid> ,<\ln\mid Re Z\mid> 
\end{equation}
are identical. In a slightly cavalier fashion we will refer to the 
quantity $\langle \ln |{\rm Re}
Z|\rangle$ as free energy, taking into account that for the case of real 
temperatures it is related simply to the usual free energy.
\begin{equation}
\label{e1}
\langle \ln |{\rm Re}
Z|\rangle=\int_{0}^{\infty}dt{\frac{e^{-t}}{t}}-\langle \int_{0}^{\infty}
dt{\frac{e^{-t|{\rm Re} Z|}}{t}}\rangle
=\int_{0}^{\infty}dt{\ln te^{-t}}+\int_{0}^{\infty}\ln td\langle
e^{-t|{\rm Re} Z|}\rangle
\end{equation}
where $\lambda=\beta \sqrt{N}$, and $\langle ...\rangle$ means averaging
by the normal distribution. Further,
\begin{equation}
\label{e2}
\langle e^{-t|{\rm Re} Z|}\rangle=\langle e^{-\phi}\rangle
=\frac{1}{\sqrt{\pi}}\int_{-\infty}^{\infty}{\it d}z
e^{-z^2/4}f(z\sqrt{t},\lambda)^{2^N}
\end{equation}
and
\begin{equation}
\label{e3}
f(A,\lambda)=\frac{1}{\sqrt{\pi}}\int_{-\infty}^{\infty}{\it d}x
e^{-x^2+i\frac{A}{2}(\exp(\lambda x)+\exp{\overline{\lambda}} x)}
\end{equation}
here we have $\beta\equiv 
\beta_1+\beta_2,\lambda\equiv\lambda_1+i\lambda_2=\sqrt{N}\beta$

We see that $|f(A,\lambda)|$ is always less than 1. When $\ln A\gg 0$ the 
$ f(A,\lambda)$
is exponentially suppressed and we get $f(A,\lambda)\approx 1$, while $\ln
A\ll U$.
To find asymptotics it is convenient to use another representation, like
[8]
\begin{equation}
\label{e4} 
f(A,\lambda)=\frac{1}{2\pi}\int_{-i\infty}^{i\infty}
{\it d}x e^{\frac{\lambda_1^2 x^2}{4}}
\int_{-\infty}^{\infty}{\it d}y  
e^{\lambda_1 xy-\frac{iA}{2}(e^{\lambda y}+e^{\bar \lambda y})}=
\end{equation}
$$=\frac{1}{2\pi}\int_{-i\infty}^{i\infty}dxe^{\frac{\lambda_1^2x^2}{4}}
\int_{0}^{\infty}{\it d}y
y^{x-1}e^{\frac{iA}{2}(y^{1+i\lambda_2/\lambda_1}+
y^{1-i\lambda_2/\lambda_1})}$$
In the formula (4) the integral is taken along the line, which passes the
point $x=0$ on the right side. To calculate asymptotic we should move the
integration line at left to catch the saddle point. There is a simple pole
at
point $0$. It is easy to check,that the corresponding residue equals $1$.
\begin{equation}
\label{e5}
f(A,\lambda)=1+\int_{-i\infty+\varepsilon}^{i\infty-\varepsilon}{\it d}x
e^{\frac{\lambda_1^2 x^2}{4}-x \ln A}g(x,A)
\end{equation} 
where
\begin{equation}
\label{e6}
g(x,A)=\int_{0}^{\infty}{\it d}y y^{x-1}e^{iy\cos \lambda_2/\lambda_1 
\ln y/A}
\end{equation} 
This representation is useful, when the saddle point $x$ belongs to the
interval $(0,-1)$.
Function $g(x,A)$ has a complicated singularity at point -1. We need in
asymptotic at $\lambda\sim\sqrt N$,$\ln A\sim N$:
\begin{equation}
\label{e7}
f(A,\lambda)-1\approx \exp[-\frac{{\ln A}^2}{\lambda_1^2}]
\end{equation} 
where we missed pre-exponents.

There are another 2 modes for $f(A,\lambda)$, besides (7). In order to
calculate 
them we are considering derivatives:
\begin{equation}
\label{e8}
\frac{d}{dA}f(A,\lambda)=\frac{i}{4\sqrt{\pi}}\int^{i\infty}_{-i\infty}{\it
d}x
e^{-x^2+\frac{iA}{2}(e^{\lambda x}+e^{\bar \lambda x})}(e^{\lambda x}+
e^{\bar \lambda x})=
\end{equation}
$$\frac{i}{4\sqrt{\pi}}e^{\frac{\lambda^2}{4}}\int_{-i\infty}^{i\infty}{\it
d}x
e^{-x^2+\frac{iA}{2}(e^{\frac{\lambda^2}{4}+\lambda x}
+e^{\frac{|\lambda|^2}{4}
+ \bar \lambda x})}+$$
$$+\frac{i}{4\sqrt{\pi}}e^{\frac{\lambda^2}{4}}\int_{-i\infty}^{i\infty}{\it
d}x
e^{-x^2+\frac{iA}{2}(e^{\frac{|\lambda|^2}{4}+\lambda x}+
e^{\frac{\bar {\lambda^2}}{4}+
\bar \lambda x})}= $$
$$=\frac{ie^{\frac{\lambda^2}{4}}}{4\pi}\int_{-i\infty}^{i\infty}{\it d}x
e^{\frac{\lambda_1^2 x^2}{4}-(\ln A+\frac{|\lambda|^2}{4})x} g(x,A)
+\frac{ie^{\frac{\bar {\lambda^2}}{4}}}{4\pi}\int_{-i\infty}^{i\infty}{\it
d}x
e^{\frac{\lambda_1^2 x^2}{4}-(\ln A+\frac{|\lambda|^2}{4})x}g(x,A)$$
Eventually we have (after shift of integration line)
\begin{equation}
\label{e9}
f'_A\approx
\frac{i(e^{\lambda^2/4}+e^{\overline\lambda^2/4})}{2}
\end{equation}
\begin{equation}
\label{e10}
f_A\approx 1+iA(e^{\lambda^2/4}+e^{\overline\lambda^2/4)}
\approx 1+iAe^{(\lambda_1^2-\lambda_2^2)/4}
\end{equation}
The similar approach gives 
\begin{equation}
\label{e11}
f''_A\approx
-\frac{e^{ {\lambda_1^2}}}{4\pi}\int_{-i\infty}^{i\infty}{\it
d}x
e^{\frac{\lambda_1^2 x^2}{4}-(\ln A+\frac{|\lambda|^2}{4})x}g(x,A)
\end{equation}
$$f(A,\lambda)\approx 1-\frac{A^2e^{\lambda_1^2}}{4}$$
So we have three phases
\begin{equation}
\label{e12}
e^{-\phi}=\frac{1}{\sqrt{2\pi}}\int_{-i\infty}^{i\infty}{\it d}z
\exp{(-z^2/4+2^N z t e^{\frac{\lambda_1^2-\lambda_2^2}{4}})}
=\exp{(-t^2 4^Ne^{\frac{\lambda_1^2-\lambda_2^2}{4}})}
\end{equation}

\begin{equation}
\label{e13}
e^{-\phi}=\frac{1}{\sqrt{\pi}}\int_{-i\infty}^{\i\infty}{\it d}z
\exp[-z^2/4-\frac{z^2 2^Nt^2e^{\lambda_1^2/2}}{2}]
\end{equation} 

\begin{equation}
\label{e14}
e^{-\phi}=\frac{1}{2\sqrt{\pi}}\int_{-i\infty}^{i\infty}{\it d}z
\exp{(-z^2/4+2^N e^{-\frac{(\ln t+\ln zu)^2}{\lambda_1^2})}}
\end{equation}

We obtain the following expressions for  free energy.
For the paramagnetic (PM) phase, without any magnetization, zeros of 
partition sum:
\begin{equation}
\label{e15}
\langle\ln |{\rm Re}Z|\rangle =N\ln 2+\frac{N(\beta_1^2-\beta_2^2)}{4}
\end{equation}
For the pin glass (SG) phase with some replica symmetry breaking:
\begin{equation}
\label{e16}
\langle\ln |{\rm Re}Z|\rangle =N\sqrt{\ln 2}\beta_1
\end{equation}
For the phase without any magnetization, but with some zeros for 
partition sum (further abbreviated as LYF; Lee-Yang
\cite{yang}, and Fisher \cite{fisher}):
\begin{equation}
\label{e17}
\langle\ln |{\rm Re}Z|\rangle =\frac{1}{2}N\ln 2+\frac{\beta_1^2}{2}N
\end{equation}
Let us consider now the diluted version of the model.
If we consider again our hamiltonian (\ref{ee1}), only with $\pm 1$ 
couplings, then chose randomly
among the all possible $C_N^p$ sets of indices ${i_1..i_p}$ some $\alpha N$
\begin{equation}
\label{ee17}
H=-\sum_{(i_1<i_2..<i_p)=1}^{\alpha N}j_{i_1..i_p}s_{i_1}..s_{i_p}
\end{equation}
 Now instead of
the normal distributions we have
\begin{equation}
\label{e18}
\rho(E)=\frac{1}{2\pi}\int_{-i\infty}^{i\infty}{\it d}E_1
e^{-EE_1-\alpha N\ln ch E_1}
\end{equation}
Here $\alpha >1$ is a ratio of couplings number to spins number.
For function $f(A,\lambda )$ we have
\begin{equation}
\label{e19}
\frac{1}{2\pi}\int_{-i\infty}^{i\infty}{\it d}E_1
e^{\alpha N\ln \cosh \beta E_1} \int_{-\infty}^{\infty}{\it d}E
e^{-\beta EE_1-\frac{iA}{2}(e^{\beta E}+e^{\bar \beta E})}=
\end{equation}
$$=\frac{1}{2\pi}\int_{-i\infty}^{i\infty}{\it d}x
e^{\alpha N \ln {\cosh {\beta x}}+x \ln A}
\int_{0}^{\infty}e^{\frac{i}{2}(y^{1+\frac{i\beta_2}{\beta_1}}
+y^{1-\frac{i\beta_2}{\beta_1}})} y^{-x}{\it d}y=$$
$$=\frac{1}{2\pi}\int_{-i\infty}^{i\infty}{\it d}x
e^{\alpha N \ln {\cosh {\beta x}}-x \ln A} g(x,A)$$
Again pole at point 0 gives
\begin{equation}
\label{e20}
f(A,\lambda)\approx1-e^{\alpha N \ln {\cosh {\beta x}}-x\ln A},
\tanh {\beta x}=\frac{\ln A}{\alpha N \beta}
\end{equation}
This asymptotics is correct, when
\begin{equation}
\label{e21}
\tanh \beta>\frac{\ln A}{\alpha N \beta}
\end{equation}
To calculate another types of asymptotics we have to consider
the derivatives $f'_A, f''_A$.
\begin{equation}
\label{e22}
f'_A=\frac{i}{4\pi}\int_{-i\infty}^{i\infty}{\it d}E
e^{-E\hat E_1 \beta+\alpha N \ln \cosh \beta E_1+\frac{iA}{2}
(e^{\beta E}+e^{\bar \beta E})} (e^{\beta E}+e^{\bar \beta E})=
\end{equation}
$$=\frac{i}{4\pi}\int_{-\infty}^{\infty}{\it d}E
e^{E\hat E_1 \beta+\alpha N \ln \cosh \beta(E_1+1)+\frac{iA}{2}
(e^{\beta E}+e^{\bar \beta E})}+$$
$$+\frac{i}{4\pi}
\int_{-i\infty}^{i\infty}{\it d}E e^{-EE_1\beta+\alpha N \ln \cosh
\beta(E_1+
\bar \beta_1/\beta)+\frac{iA}{2}(e^{\beta E}+e^{\bar \beta E})}=$$
$$=\frac{i}{4\pi}\int_{-i\infty}^{i\infty}e^{\alpha N \ln \cosh
{\beta(x+1)}+x\ln A } g(x,A){\it d}x+$$
$$+\frac{i}{4\pi}\int_{-i\infty}^{i\infty}e^{\alpha N \ln \cosh
\beta(x+\bar \beta/\beta)+x\ln A} g(x,A){\it d}x$$
We have to move integration line left to catch the sadle point. Again we
intersect the pole at point 0.
\begin{equation}
\label{e23}
f(A,\lambda)\approx1+\frac{iA}{2}(e^{\alpha N \ln \cosh \beta}+e^{\alpha N
\ln \cosh \bar \beta})
\end{equation}
This asymptotics is correct if for a solution
\begin{equation}
\label{e24}
\frac{\ln A}{\alpha N \beta}=\tanh(\bar \beta+\beta x)
\end{equation}
we have the constraint $Re x<0$.

Let us consider now the third type of asymptotics. The consideration of
second derivative reads:
\begin{equation}
\label{e25}
\frac{d^2f}{dA^2}=-\frac{1}{8\pi}\int_{-i\infty}^{i\infty}{\it d}E
e^{-EE_1\beta+\alpha N \ln \cosh{\beta E_1}+\frac{iA}{2}(e^{\beta E}+
e^{\bar \beta E})}
(e^{2\beta E}+e^{2\bar \beta E}+2e^{2\beta_1 E})=
\end{equation}
$$=-\frac{1}{4\pi}\int_{-i\infty}^{i\infty}e^{\alpha N\ln \cosh
(x+2\beta_1)-E_1\ln A} g(x){\it d}x$$
$$-\frac{1}{8\pi}\int_{-i\infty}^{i\infty}e^{\alpha N\ln \cosh
(x+2\beta )-E_1\ln A} g(x){\it d}x$$
$$-\frac{1}{8\pi}\int_{-i\infty}^{i\infty}e^{\alpha N\ln \cosh
(x+2\bar \beta )-E_1\ln A} g(x)$$
The shift of integration line  intersects a pole, and
gives
\begin{equation}
\label{e26}
\frac{d^2 f}{dA^2}=-\frac{1}{8\pi}[2e^{\alpha N\ln \cosh
(2\beta_1)}-e^{\alpha N\ln (\beta E_1+2\beta_1)-E_1\ln A}]
\end{equation}
where
\begin{equation}
\label{e27}
\alpha N\beta \tanh(\beta E_1+2\beta_1)=-\ln A
\end{equation}
\begin{equation}
\label{e28}
A-1\approx-\frac{1}{4\pi}e^{\alpha N \ln \cosh(2\beta_1)}
\end{equation}
Again we have 3 mods.
\begin{equation}
\label{e29}
e^{-\phi}=\frac{1}{2\sqrt{\pi}}\int_{-i\infty}^{i\infty}e^{-z^2/4+zi2^Nt
[e^{\alpha N\ln \cosh (\beta)}+e^{\alpha N\ln \cos (\bar \beta )}]}
\end{equation}

\begin{equation}
\label{e30}
e^{-\phi}=\frac{1}{2\sqrt{\pi}}\int_{-i\infty}^{i\infty}\exp{[-z^2/4-2^Ne^{-\frac
{(\ln t+\ln z)^2}{\lambda_1^2}}]}
\end{equation}

\begin{equation}
\label{e31}
e^{-\phi}=\frac{1}{2\sqrt{\pi}}\int_{-i\infty}^{i\infty}\exp{(-z^2/4-z^2
2^4N t^2
e^{\alpha N\ln \cosh 2\beta_1})}
\end{equation}

Eventually we have three expressions for free energy.
For the para magnetic phase:
\begin{equation}
\label{e32}
<\ln |Re z|>=N\ln 2+\alpha N Re \ln \cosh \beta
\end{equation}
For the LYF phase:
\begin{equation}
\label{e33}
<\ln |Re z|>=\frac{N}{2}\ln 2+\frac{\alpha N}{2}\ln \cosh 2\beta_1
\end{equation}
For the SG phase:
\begin{equation}
\label{e34}
<\ln Re|z|>=\alpha N \beta_1 \tanh \beta_c
\end{equation}
where for the $\beta_c$ we have
\begin{equation}
\label{e35}
\alpha [\beta_c \tanh \beta_c-\ln \cosh \beta_c]= \ln 2
\end{equation}
The search of other possible phases like
\begin{equation}
\label{e36}
\frac{\alpha N}{4}\ln \cosh 4\beta_1+\frac{N}{4}\ln 2
\end{equation}
gives the same equations, as for the case of real $T$. So we
have not any new phase, besides those obtained in Eqs.  
(\ref{e32}-\ref{e34}).
The boundary lines between 2 phases one can be found from the
coincidence of 2 equations .

The boundary between LYF and SG phase is the line
\begin{equation}
\label{e37}
\beta_1=\beta_0, \beta_0<\beta_2<\infty
\end{equation}
where the multicriticity point $(\beta_1,\beta_0)$ lives on the
intersection
of line (\ref{e37}) with the boundary SG, LYF.
\begin{equation}
\label{e38}
\sin^2{\beta_2}=\frac{2^{1/\alpha}-1-\tanh^2{\beta_1}}{1-\tanh^2{\beta_1}}
\end{equation}
For the case of ferromagnetic couplings we have the ferromagnetic
phase with the old expression with $\beta_1\not=0, \beta_2=0$.
Let us consider the density of partition zeros near the $\beta_1$ axes.
For the density of zeros we have [9]expresion
\begin{equation}
\label{e41}
\frac{1}{2\pi}(\frac{d^2}{d\beta_1^2}+\frac{d^2}{d\beta_1^2})< \ln |Z|>
\end{equation}
For the LY phase we have
\begin{equation}
\label{e42}
\rho(\beta_1,\beta_2)=\frac{1}{\pi}\frac{\alpha}{\cosh(\beta_1)^2}
\end{equation}
Near the PM-SG transition line we obtain:
\begin{equation}
\label{e43}
\rho(\beta_1,\beta_2)=\frac{1}{\pi}\frac{\alpha}{\cosh(\beta_1)^2}
\delta(\beta_1+\beta_2-\beta_c)(\beta_1-\beta_c)
\end{equation}

Now we consider the phase structure of Generalized Random
Energy Model in the domain of complex $T$.
The SG phase transition conditions are the old ones, with the $\beta_1$ 
instead of $1/T$.

The conditions for phase transition to 3-rd (LYF) phase is more
complicated. We have to consider an expression like
\begin{equation}
\label{e39}
<\prod_{\alpha}\prod_{\beta_{\alpha}}
\exp i\frac{A}{2}[e^{\lambda_\alpha E_{\alpha}+\lambda_\beta E_{\beta}}+
e^{\overline \lambda_\alpha E_{\alpha}+\overline \lambda_\beta E_{\beta}}]>
\end{equation}
while expanding the exponent (to get LYF phase), we found  
\begin{equation}
\label{e40}
<\prod_{\alpha}\prod_{\beta_{\alpha}}
1-A^2/4[e^{(\lambda_\alpha+\overline
\lambda_\alpha) E_{\alpha}+(\lambda_\beta +\overline \lambda_\beta) 
E_{\beta}}]>
\end{equation}
This expression showes, that if at some level of hierarchy exist LY
phase, then
higher could not exist PM, so stays only the SG one.

Let us consider the diluted version of GREM [13], with infinite number of
hierarchy $M$.
We can consider the case of large $M$ with smooth distribution of 
$z_k$(number of couplings at level k) and $N_k$ (number of k-th level
branches is $2^{N_k}$). In this case we can introduce continuous variable
 $v=\frac{k}{M}$
 between
$0$ and $1$, labeling the level of hierarchy
and define distributions 

\begin{eqnarray}
\label{l44}
 z_k\equiv {\rm d}z=z{\rm d}v,\qquad  N_k\equiv {\rm d}N=n'(v){\rm d}v
\qquad{\rm d}{v}=\frac{1}{M}
\end{eqnarray}
where $n(v)$ is a given function (entropy in bits-s). The variable 
$v$ ( $0<v<1$) 
parameterizes the level of the hierarchical tree and $z$ is just a
parameter 
(for our spin system
z is a total number of couplings and parameter $v$ labeling the level of 
hierarchy). 

Similarly to the case of
dilute GREM at real T we found:
\begin{eqnarray}
\label{e45} 
-\frac{\beta F}{N}=& & z(1-v_2(\beta))Re \ln \cosh \beta
+n(v_2(\beta))\ln 2+n(v_2(\beta))\ln 2/2\\
& &+z(v_2(\beta)-v_1(\beta))/2\ln \cosh 2\beta_1
+ z \beta_1\int_{0}^{v_1(\beta)}{\rm d}v_0 g(\frac{z}{n'(v_0)})\nonumber
\end{eqnarray}
where
$v_2(\beta),v_1(\beta)$
are defined from the equations
\begin{equation}
\label{e46}
2zRe \ln \cosh \beta+n'(v_2(\beta))\ln 2=z\ln \cosh 2\beta_1
\end{equation}
$$zRe \ln \cosh \beta+n'(v_1(\beta))\ln 2=\beta_1 zg(z/n'(v_1))$$
 and function $g(x)$from the
\begin{equation}
\label{e47}
\frac{1}{2}(1+g)\ln (1+g)+\frac{1}{2}(1-g)\ln(1-g)]
=\frac{\ln 2}{x}
\end{equation}
For the case of Edwards-Anderson model placed on d-dimensional hypercubic
lattice
\begin{equation}
\label{e48}
z=Nd,\quad E=-vNd \quad n(v)=\frac{Ns(-vdN)}{\ln 2}
\end{equation}
here $E$ is energy, and $s(E)$ is entropy as function of the energy for
corresponding ferromagnetic Ising model.
At given   $\tilde \beta_0$ we can define corresponding $v_0$
 as minus energy per bond for  ferromagnetic
 model at temperature $\frac{1}{\tilde \beta_0}$.
\begin{equation}
\label{e49}
v_0(\tilde\beta_0)=-\frac{E(\tilde\beta_0)}{Nd}
\end{equation} 

We can remember from thee definition of temperature 
\begin{equation}
\label{e50}
\frac{{\rm d} s}{{\rm d } E}=\frac{1}{\tau}\equiv \tilde\beta
\end{equation}
We obtain for the free energy
\begin{equation}
\label{e51}
-\frac{\beta F}{Nd}=(1-v_2(\beta))Re\ln \cosh \beta+s(v_2(\beta))+
(v_2(\beta)-v_1(\beta))/2\ln \cosh 2\beta_1
\end{equation}
$$+s(v_2(\beta))/(2Nd)
+\beta_1\int_{0}^{v_1(\beta)}{\rm d}v_0
f(\frac{\ln 2}{\tilde\beta(v_0)})$$

Integrating by parts in the last term we get
\begin{equation}
\label{e52}
-\frac{\beta F}{Nd}=(1-v_2(\beta))Re\ln \cosh \beta+s(v_2(\beta))
(v_2(\beta)-v_1(\beta))/2\ln \cosh 2\beta_1
\end{equation}
$$+(s(v_2(\beta)-s(v_1(\beta)))/(2Nd)
-\beta_1 \int_{0}^{\tilde \beta_1}{\rm d}
\tilde \beta_0 \frac{2v_0(\tilde\beta_0)}{\ln\frac{1+y}{1-y}}
+v_1(\beta)\beta_1 y(\tilde\beta_1)$$
where $y$ as a function of $\tilde\beta_0$  is defined from the equation
\begin{equation}
\label{e53}
y=g(\frac{\ln2}{\tilde \beta_0})
\end{equation}
function $v_0(\tilde\beta)$ is defined from the (\ref{e49}).
 and $\tilde\beta_1(\beta)$ is defined from the equation
\begin{equation}
\label{e54}
v_1(\beta)
=-E(\tilde \beta_1)/(Nd)
\end{equation}
where $v_1(\beta)$ is defined from the (\ref{e46}).
Here  $E(\tilde \beta_1)$ is the energy of ferromagnetic model
on the same lattice.

We propose the method for calculating approximate statistical mechanics of
disordered models on finite
lattices. It allows us also to identify Lee-Yang-Fisher zeroes. In the 
case of real temperatures
our generalization of REM gave 5\% accuracy. If such an accuracy will be
available here, then this method could be very efficient. 

We could succeed due to a simple observation (\ref{e40}) that on hierachy 
of generalized REM new 
(Lee-Yang-Fisher) phase could exist only on levels between paramagnetic
and spin glass. On the other hand, there is no replica symmetry breaking.
It is very interesting to consider the similar issues in strings where 
the ideas and methods of spin-glass theory have been applied recently. 

We thank Fundacion Andes grant-C-13413/1 for a partial support.

\end{document}